\documentclass[twocolumn,twoside,slac_two]{revtex4}
\usepackage{graphicx}
\usepackage{fancyhdr}
\usepackage{graphics}
\usepackage{epstopdf}
\usepackage{textpos}
\begin{document}

%%%%%%%%%%%%%%%%%%%%%% WRITE THE TITLE HERE %%%%%%%%%%%%%%%%%%%
\title{\centering ATLAS Jet Energy Scale}
%%%%%%%%%%%%%%%%%%%%%% WRITE THE AUTHOR HERE %%%%%%%%%%%%%%%%%

%%% Please insert your personal picture here!

\author{
\centering
\begin{center}
D. Schouten$^{2}$, A. Tanasijczuk$^{1}$, M. Vetterli$^{1,2}$, on
behalf of the ATLAS Collaboration
\end{center}}
\affiliation{\centering $^{1}$Simon Fraser University and $^{2}$TRIUMF, Canada}
%%%%%%%%%%%%%%%%%%%%%% WRITE THE ABSTRACT HERE %%%%%%%%%%%%%%%%
\begin{abstract}
Jets originating from the fragmentation of quarks and gluons are the most common, and complicated,
final state objects produced at hadron colliders. A precise knowledge of their energy calibration 
is therefore of great importance at experiments at the Large Hadron Collider at CERN, while is 
very difficult to ascertain. We present in-situ techniques and results for the jet energy 
scale at ATLAS using recent collision data. ATLAS has demonstrated an understanding of the necessary 
jet energy corrections to within $\approx$ 4\% in the central region of the calorimeter.
\end{abstract}

\maketitle
\thispagestyle{fancy}

% References should be done using the  \cite, \ref, and \label commands

\section{Introduction}
The ATLAS experiment has been collecting collision data from the Large
Hadron Coller (LHC) since early 2010. Currently, the ATLAS jet
calibration is derived from Monte Carlo simulations, while its
associated uncertainty is derived from a combination of single hadron
and dijet response measurements, and systematic variations in Monte
Carlo simulations. 
In order to validate this approach, ATLAS has employed a number of 
approaches to demonstrate an understanding of the jet 
energy scale.

In Section 1 the single particle response measurement in the central
barrel region is presented. This is extrapolated to the endcap
regions of the calorimeter using the diject relative response
measurement, as described in Section 2. Section 3 details the photon +
jet measurements. Finally, in Section 4 a summary of all of the
validation methods for the ATLAS jet energy scale (JES) is presented.

\section{Single Particle Response}
\begin{figure}
  \centering
  \includegraphics[width=0.3\textwidth]{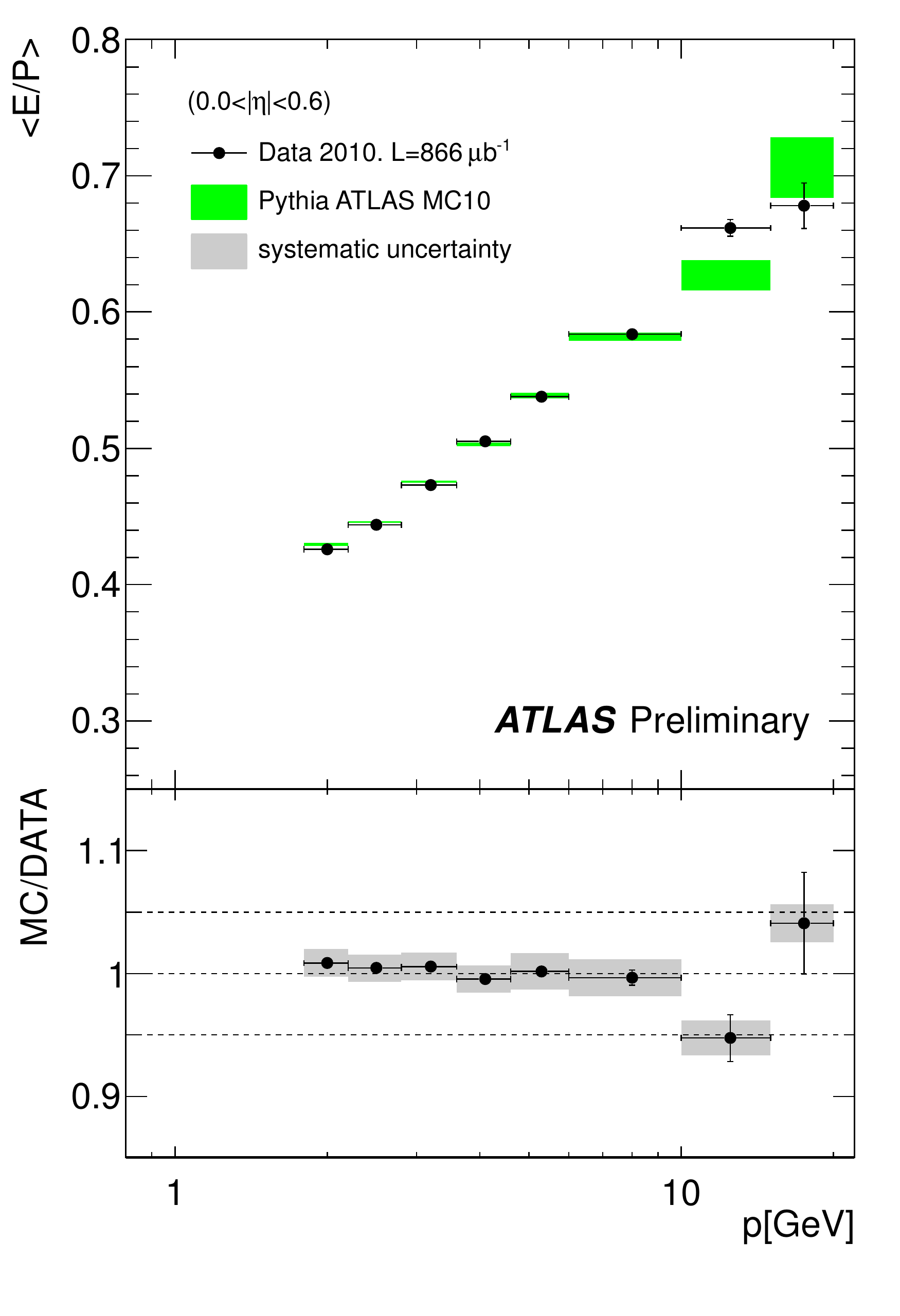}
  \caption{Comparison of $E/p$ measured in data (black points) and
    Monte Carlo simulations (green blocks) \cite{EoverP}. The ratio of data to MC is
  also shown, in which the grey blocks represent the systematic and
  the bars represent the statistical uncertainty.}
  \label{fig:EoverP}
\end{figure}

The basic idea underyling the single particle response is to measure
the calorimeter response for isolated single 
particles by comparing the energy and momentum
(tracking) measurements, namely $E/p$, under the assumption that the
tracking measurement is very precise. Uncertainties for single particles are derived from
deviations of this measurement in simulations compared to data. Then, these
are extrapolated to jet uncertainties using simulations. 
Although the translation from single particles
to the jet context is non-trivial, it has been exhaustively
cross-checked and is found to have small uncertainty. For charged
particles in the momentum range $0.5 < p < 20$ 
GeV the $E/p$ measured in situ is used to determine the response
 \cite{EoverP}. 
%% In order to account for the neutral background component to the measured
%% $E/p$, the background is estimated by looking in an annulus $0.1 <
%% \Delta R = \sqrt{\Delta\eta^{2} + \Delta\phi^{2}} < 0.2$ around
%% minimum ionizing particles in the EM calorimeter. This contribution is
%% then subtracted from the measured $E/p$ ratio. 
A comparison of $E/p$ in data to that in Monte Carlo is shown in Figure \ref{fig:EoverP}.

\section{Relative Response}
\begin{figure}
  \centering
  \includegraphics[width=0.4\textwidth]{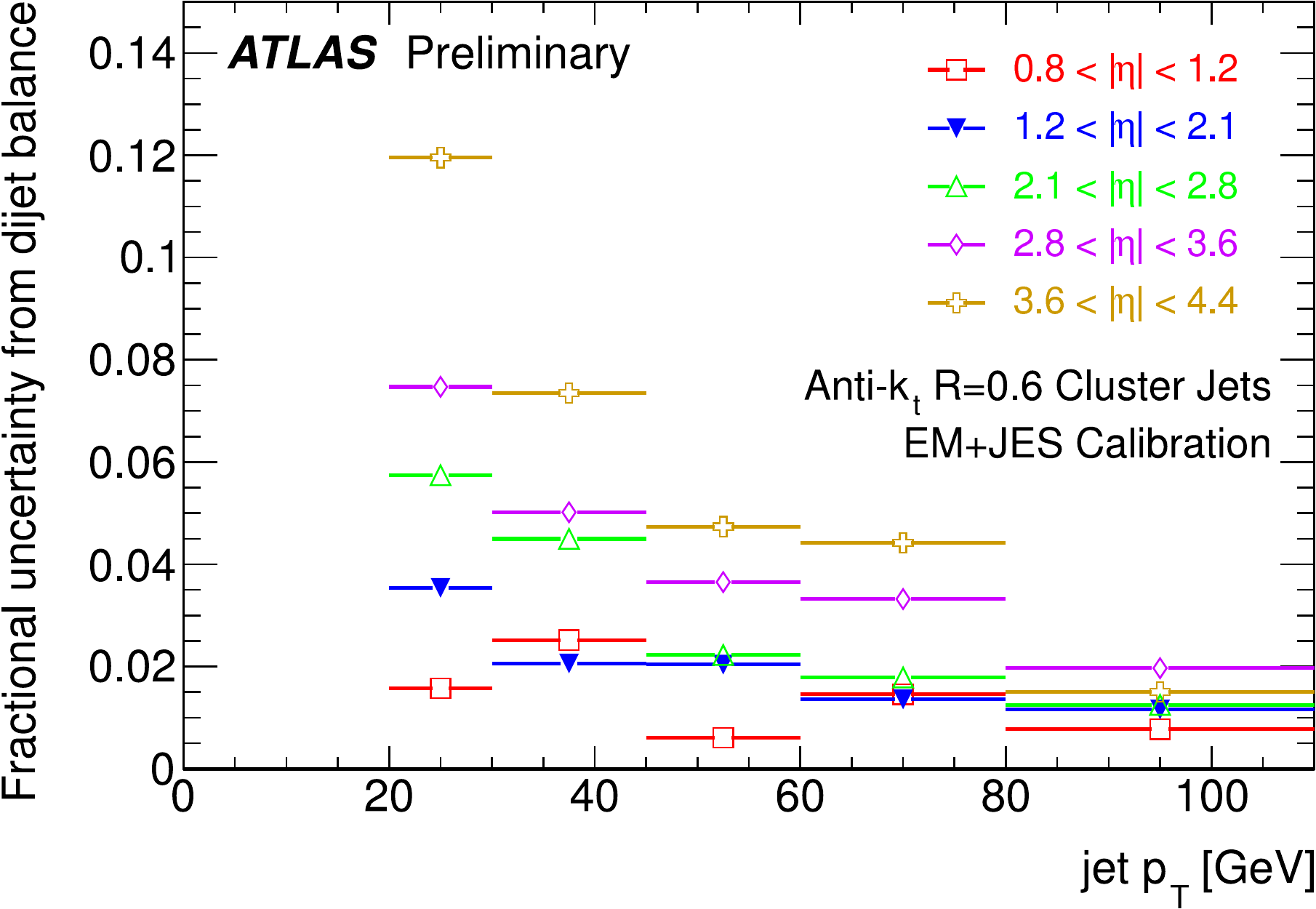}
  \caption{Systematic uncertainty derived for the relative correction
    using dijet balance as a function of jet $p_{T}$ for various
    regions of pseudorapidity \cite{InterCalib}.}
  \label{fig:InterCalib}
\end{figure}

For jets outside of the central barrel, the
response for the central region is extrapolated using a dijet balance
technique  \cite{InterCalib}. This procedure measures a response for
a jet relative to the central region under the assumption of momentum
balance of the dijet system, and compares it to the result in
simulations. The JES uncertainty in the endcap region
is then the sum in quadrature of the uncertainties in the central region
and the dijet relative response measurements.
Currently, the latter component (shown in Figure
\ref{fig:InterCalib}) is the dominant one in the forward region,
due to a disagreement  between different Monte Carlo generators in
the modelling of the reference balance assumption.

\section{Response from Photon + Jet Events}
\begin{figure}[t!]
  \centering
  \includegraphics[width=0.4\textwidth]{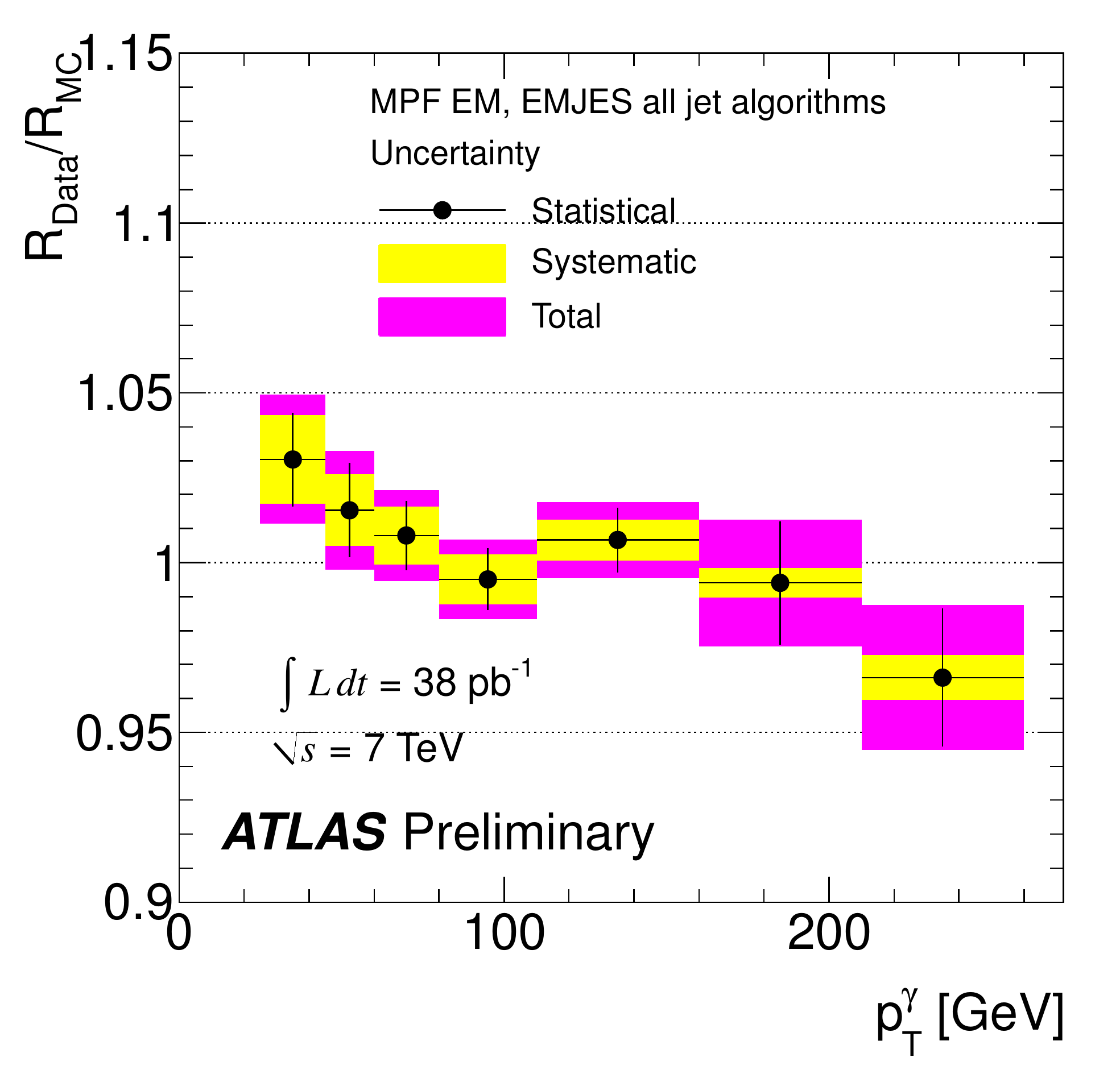}
  \caption{Ratio of $MPF_{DATA} / MPF_{MC}$ versus photon
    $p_{T}$ \cite{GammaJet}. The total uncertainty is indicated by the
    magenta band, while the systematic uncertainty only is shown by
    the yellow band.} 
  \label{fig:GammaJet}
\end{figure}

Because photons are well-measured objects, one can
directly measure the jet response by using the principle
of momentum balance between a photon and recoil jet in
photon + jet events. One technique, known as the
missing ET projection fraction (MPF) directly measures
the total calorimeter response to jets by balancing the
hadronic recoil against the photon. The MPF equation is:
\begin{equation}
  R_{MPF} = 1 + \frac{E_{T}^{miss}\cdot
    \hat{n_{\gamma}}}{p_{T}^{\gamma}}.
\end{equation}
Directly balancing the photon and jet in these events is a
complementary technique, and is differently sensitive to radiative
effects \cite{GammaJet}. 
A comparison of the MPF in data and simulation is shown in Figure
\ref{fig:GammaJet}. The Monte Carlo simulation agrees with the data to
within a few percent over the entire range of photon $p_{T}$. 

\section{Summary of Jet Energy Scale}
Besides the techniques summarized above, comparing track and
calorimeter jets and also measuring transverse momentum balance in
multi-jet final states are useful probes of the jet energy scale. 

The track jet comparison test works off of the assumption that the ratio of the
charge particle momentum to the total jet momentum is tightly
constrained. Thus, by directly measuring the ratio of the track jet to
the matched calorimeter jet momentum in data and in simulations, the validity
of the simulation can be determined \cite{TrackJet}. 

Employing momentum balance in multi-jet final states in which a high
$p_{T}$ jet recoils against many lower $p_{T}$ jets allows for
validation at very high $p_{T}$ \cite{MultiJet}. This is because the
uncertainties can be ascertained for the recoil jets using the
standard approaches described above, since they are in a reachable
$p_{T}$ range for these methods. Then the momentum balance of the
recoil system and lead jet can be compared in data and simulations.

\begin{figure}[h!]
  \centering
  \includegraphics[width=0.45\textwidth]{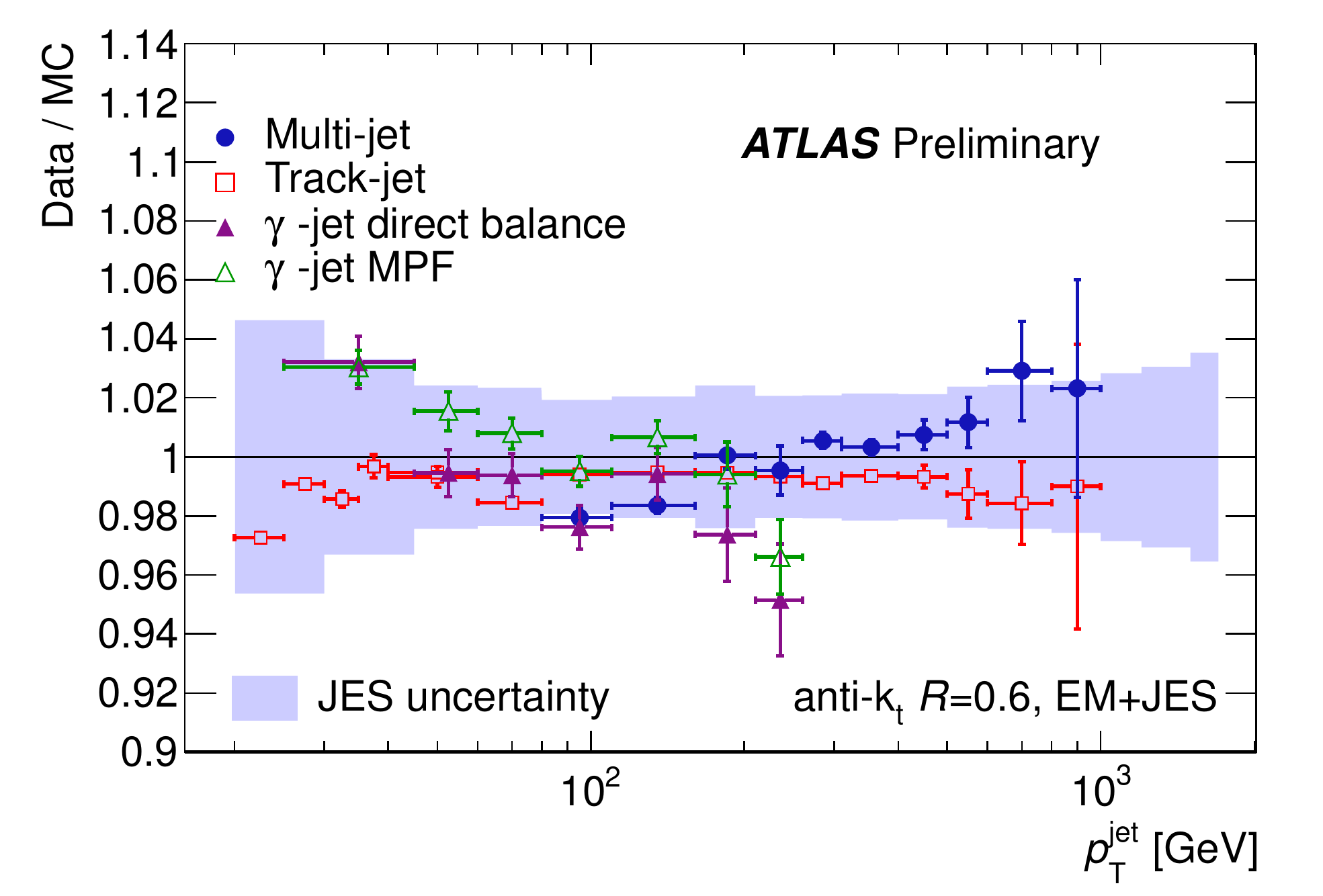}
  \caption{Summary of all the jet energy scale measurements performed
    by ATLAS, for the central calorimeter region  \cite{JES}. The
    shaded graph is the uncertainty for the Monte Carlo
    simulation-based jet calibration.}
  \label{fig:JES}
\end{figure}

By combining these techniques with those described
above, a robust validation of the JES and its uncertainty can be shown
\cite{JES}. This is summarized in the central region in Figure \ref{fig:JES}.  

In summary, the jet energy scale and its uncertainty in ATLAS, 
derived from Monte Carlo simulations, has been extensively validated
to within $\pm 4\%$ in the central region of the calorimeter for $p_T
> 20$ GeV (or $\pm 2.5\%$ for $p_T > 40$ GeV) using a
variety of complementary approaches.  

\bigskip % extra skip inserted

%% Create the reference section using BibTeX:
\bibliography{basename of .bib file}

\end{document}